\documentclass[aps,pre,onecolumn,floatfix,natbib=false]{revtex4-2}

\usepackage[english]{babel}
\usepackage{listings}
\usepackage{tikz}
\usepackage{subcaption}
\usepackage{float}
\usepackage{comment}
\usepackage{minted}
\usepackage{xcolor}
\usepackage{fvextra}
\usepackage{amsmath}
\usepackage{graphicx}
\usepackage{physics}
\usepackage[colorlinks=true, allcolors=blue]{hyperref}
\usepackage[letterpaper,top=2cm,bottom=2cm,left=3cm,right=3cm,marginparwidth=1.75cm]{geometry}

\definecolor{bg}{rgb}{0.12, 0.12, 0.12}
\definecolor{LightGray}{gray}{0.9}

\usetikzlibrary{positioning, arrows.meta, shapes.misc, shapes.geometric, calc, fit}

\tikzset{block/.style={rectangle, draw, fill=blue!20, minimum height=3em, minimum width=6em, text centered, text width=5.5em}, arrow/.style={-Latex}, line/.style={draw, -latex'}}

\begin{document}

\title{Transpiling quantum circuits by a transformers-based algorithm}
\author{Michele Banfi$^{1}$, Paolo Zentilini$^{1,2}$, Sebastiano Corli$^{1}$, Enrico Prati$^{1,2}$}

\email{enrico.prati@unimi.it}

\affiliation{$^1$Department of Physics Aldo Pontremoli, Università degli Studi di Milano, Italy}
\affiliation{$^2$Istituto di Fotonica e Nanotecnologie, Consiglio Nazionale delle Ricerche, Italy}

\begin{abstract}
    Transformers have gained popularity in machine learning due to their application in the field of natural language processing. They manipulate and process text efficiently, capturing long-range dependencies among data and performing the next word prediction. On the other hand, gate-based quantum computing is based on controlling the register of qubits in the quantum hardware by applying a sequence of gates, a process which can be interpreted as a low level text programming language.
    We develop a transformer model capable of transpiling quantum circuits from the qasm standard to other sets of gates native suited for a specific target quantum hardware, in our case the set for the trapped-ion quantum computers of IonQ. The feasibility of a translation up to five qubits is demonstrated with a percentage of correctly transpiled target circuits equal or superior to 99.98\%. Regardless the depth of the register and the number of gates applied, we prove that the complexity of the transformer model scales, in the worst case scenario, with a polynomial trend by increasing the depth of the register and the length of the circuit, allowing models with a higher number of parameters to be efficiently trained on HPC infrastructures.
\end{abstract}

\maketitle

\section{Introduction}

A transformer capable of accurately translating quantum circuits between IBM and IonQ gate sets, achieving over $99.98\%$ accuracy with scalable complexity, is developed.
Quantum circuits are ordered sequences of logic gates acting on qubits. The most conventional way to implement such gates is via unitary operators, even though another feasible approach, in quantum mechanics, is through disruptive measurements (the so called one-way, or measurement-based quantum computing)~\cite{corli2025quantum,browne2016one,barenco1995elementary}.
Since current devices operate in the noisy intermediate-scale quantum (NISQ) regime, the practical execution of a circuit depends crucially on its efficiency. 
Fewer qubits and shorter gate sequences reduce the accumulation of hardware noise and increase the fidelity of the computation.
The process of transforming a circuit into a form that is both efficient and physically executable is known as quantum compiling \cite{maronese2022quantum,khaneja2001cartan,corli2025gauge}.
Quantum compiling involves transformations that preserve the logical behavior of a quantum algorithm while improving or adapting its circuit representation.
Such process can reduce the depth of the circuit, break down complex operations, or restructure the circuit to meet the physical constraints of a specific hardware platform. 
In fact, each quantum technology (superconducting qubits~\cite{arute2019quantum}, trapped ions~\cite{haffner2008quantum,blatt2012quantum}, neutral atoms~\cite{henriet2020quantum}, spin in semiconductors~\cite{de2023silicon} or photonic systems~\cite{o2007optical,takeda2017universal}) offers a distinct set of native gates determined by the underlying physical interactions that it can directly implement. 
Therefore, a circuit written for one platform generally can not be executed on another without being adapted and re-expressed. 
Such translation, known as quantum transpiling~\cite{hua2023qasmtrans,kamaka2020quantum,wesley2024linguaquanta}, converts a circuit from one set of gates to an equivalent sequence compatible with the target hardware and its native gates. 
As in natural language translation, the structure of the sequence can change significantly, but its meaning, i.e., the quantum algorithm encoded by the logic circuit, must remain exactly the same.


Use of transformer models in quantum computing workflows has been recently proposed \cite{russo2025optimizing, ruiz2025quantum}. 
Early applications include their use in variational quantum eigensolvers (VQE), where transformers help identify ground state configurations of parametrized quantum circuits~\cite{nakaji2024generative}.
They have also been employed to study and predict the expressiveness of quantum \textit{ansatze}, providing insight into how circuit architectures explore the underlying Hilbert space~\cite{zhang2025learning}.
Beyond heuristic or variational approaches, models such as Ket-GPT have demonstrated the ability to generate complete sequences of quantum logic gates, illustrating the potential of large language models to operate directly on circuit-level representations \cite{apak2024ketgpt}.
Some of us has previously studied data-driven strategies for quantum compiling through complementary approaches, including evolutionary techniques based on genetic algorithms \cite{de2015universal}, reinforcement learning frameworks capable of autonomously optimizing circuit structures \cite{moro2021quantum,semola2022deep}, and gauge theory methods designed to apply structural constraints and improve compilation efficiency \cite{corli2024measurement,corli2025gauge2}. 
Here, instead, we focus on the problem of translating circuits between the native gate sets of different hardware platforms using transformer architecture. 
In this work, we demonstrate that a transformer model is capable of accurately translating quantum circuits of up to five qubits between distinct vocabularies of quantum gates, achieving an accuracy rate that consistently exceeds 99.98\%. 
The model learns to convert circuits expressed in the native gate set of the IBM hardware to the native gate set used by the trapped-ion quantum processors of IonQ.
To enable the translation, we represent each circuit using OpenQASM, the hardware-independent quantum assembly language supported by Qiskit and widely adopted as the textual standard for describing quantum circuits. 
The transformer operates directly on this textual representation, taking the OpenQASM description of a source circuit and generating the corresponding sequence of gates compatible with the target device. 
The model effectively performs circuit-level translation in a manner analogous to natural language translation, but within the structured symbolic domain of quantum gate operations.

The structure of this work is as follows. 
Section \ref{sec:methods} describes the methodology used to develop the model, and the two different approaches to transpile from IBM circuits to the IonQ ones, i.e. starting from a circuit composed by a set of parametric gates (rotations) or rather by a universal set decomposed through the Solovay-Kitaev algorithm. Section \ref{sec:results} presents the experiments and results. Section \ref{sec:conclusions} consists of the conclusions and perspectives on future developments.

\begin{figure}[H]
    \centering
    \includegraphics[width=0.95\textwidth]{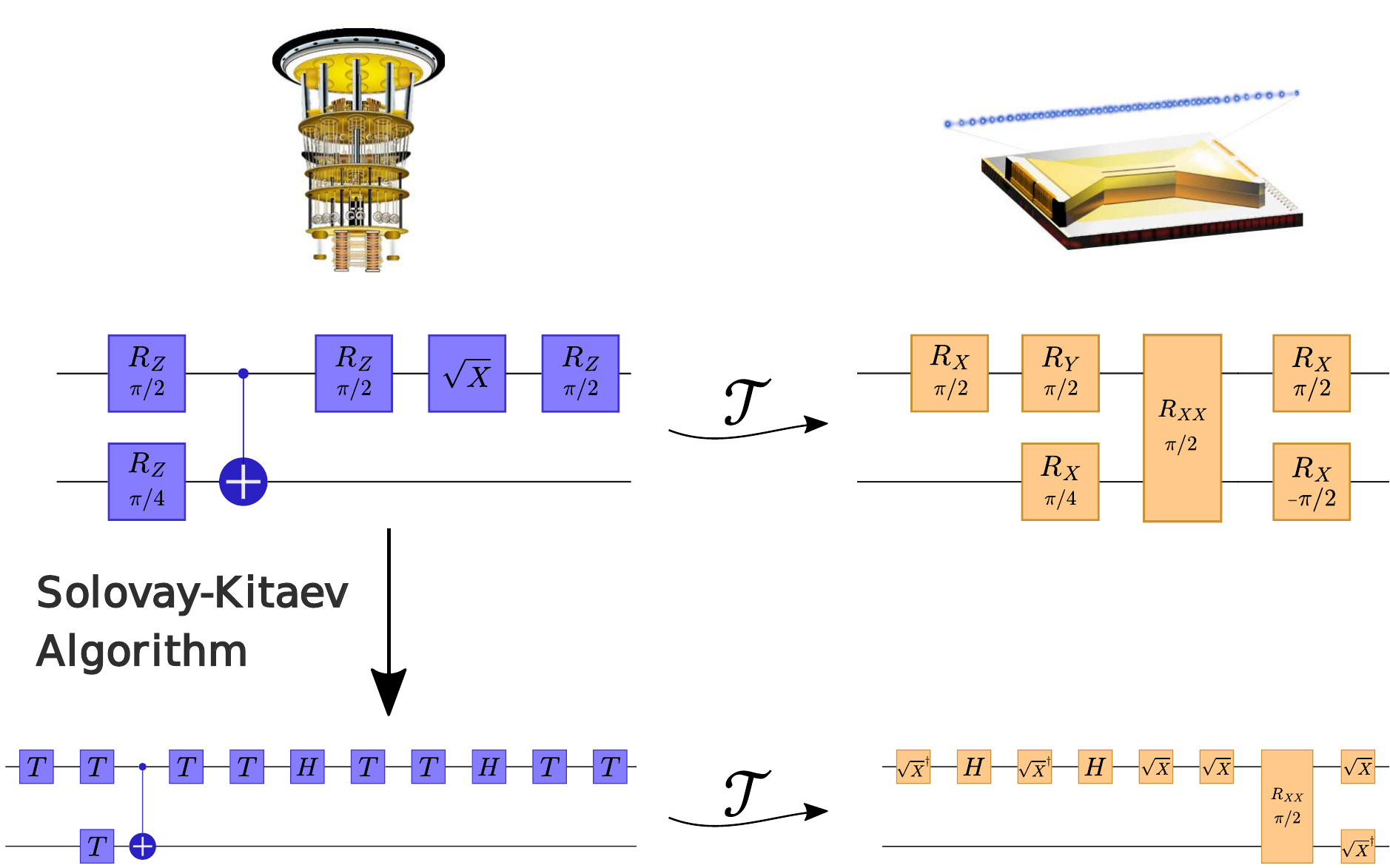}
    \caption{The input feeding the transformer is provided as a circuit composed by IBM native gates (in the picture, the violet circuit on the top-left). The same circuit is then transpiled by the transformers $\mathcal{T}$ into its IonQ counterpart (orange circuit) engaging the native gates from the trapped ions platform. The same original IBM circuit is thus decomposed through the Solovay-Kitaev algorithm into the universal set of native gates employed by IBM backends,  $\{\hat T, \hat T^\dagger, \hat H \}$, then transpiled by the transformers $\mathcal{T}$ into a universal native set of gates suitable for IonQ, i.e. $\{\sqrt{\hat X}, \sqrt{\hat X}^\dagger, \hat H \}$.}
    \label{fig:overallView}
\end{figure}

\section{\label{sec:methods}Implementation of the transformer-based transpiler}
Transformer architecture has recently become the leading model for sequential data processing in machine learning. 
Using stacked encoder and decoder blocks, transformer architecture maps an input sequence to a high-dimensional latent representation and generates the corresponding output sequence~\cite{lin2022survey}. 
Through self-attention mechanisms, transformers capture long-range dependencies by projecting tokens into a vector space where semantic and structural relationships are encoded.
The encoder identifies the relevant features of the input, while the decoder uses such information to produce a coherent and contextually appropriate output.

The purpose of this Section is to describe how to recast the transformer architecture to quantum transpiling purposes and the training of the model. The general picture of the methods presented in this work is framed in Figure \ref{fig:overallView}, while the details about the transpiling across equivalent circuits via the transformers are depicted in Figure \ref{fig:pipeline}.
We detail the preparation of the dataset, the design of the tokenizer to encode quantum circuits into a machine learning suitable data, and the architecture of the transformer model employed for sequence translation. 
Finally, we present the loss function adopted during training, and we discuss the role of physics informed components in guiding the model towards predictions. A detailed explanation about how the transformers work is provided in Appendix \ref{sec:appendix_transformer}.

\subsection{Pipeline of the model}

\begin{figure}[H]
    \centering
    \includegraphics[scale=0.5]{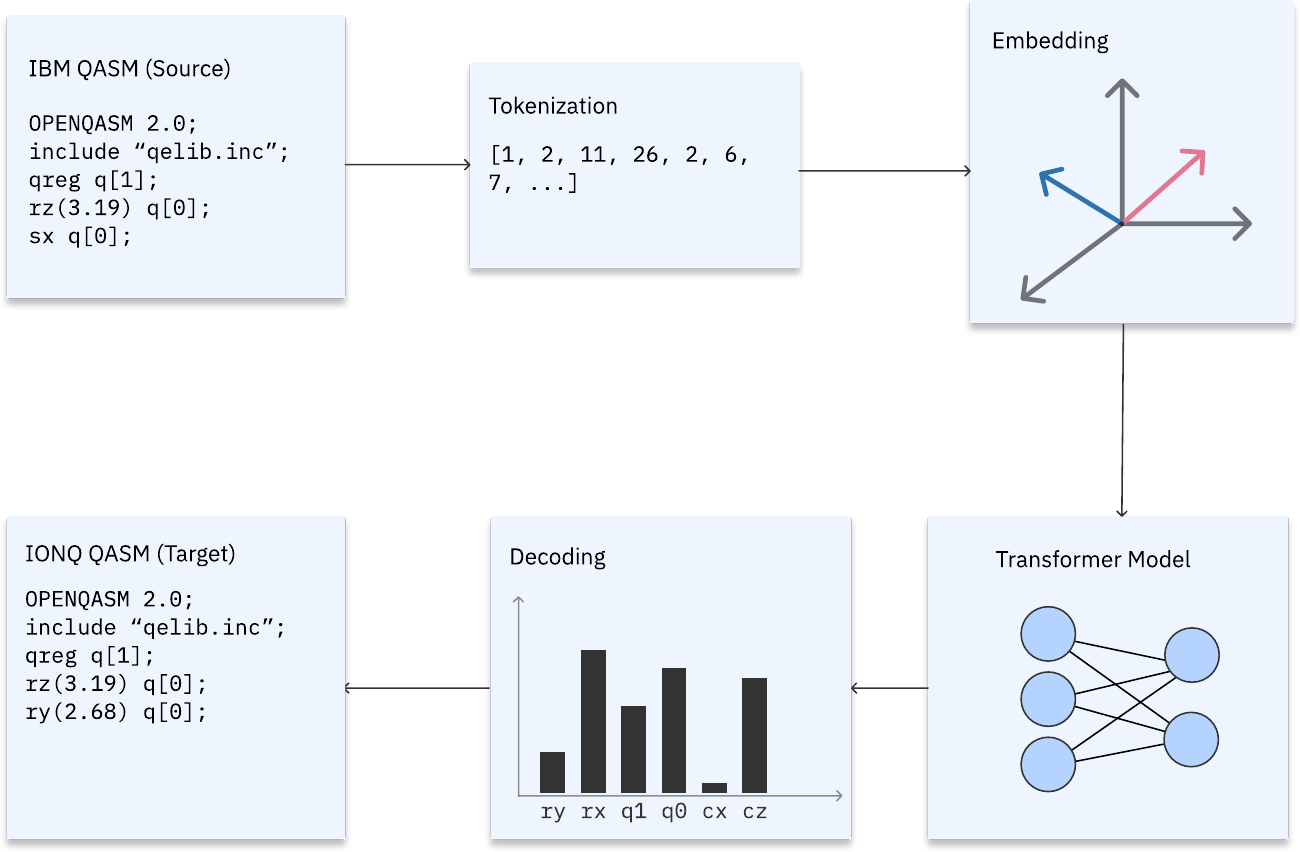}
    \caption{Pipeline of the proposed architecture. The process refers to the steps composing the $\mathcal{T}$ transformation depicted in Figure \ref{fig:overallView}. Starting from the QASM source code (here based on the IBM Eagle backend), the tokenization projects into the embedding latent Euclidean space. Thereafter, the input (i.e. the tokens) is processed throughout the transformer model and eventually decoded into the target QASM (in the following, IonQ). 
}
    \label{fig:pipeline}
\end{figure}

As first step, the preparation of the data to train the transformer is detailed. The overview of the pipeline is shown in Figure \ref{fig:pipeline}, while the description of each single component is discussed in the following Subsections.

%
\subsubsection{Data Preparation}

The training dataset is composed by pairs of representations of the same logical-circuit in the respective IBM and IonQ gate sets.
These pairs of circuits are randomly generated through already defined functions implemented in Qiskit, whose snippet code is shown in Appendix \ref{app:snippet_example}. In order to help the model generalize better, varying parameters are passed as arguments to the function. Among these parameters, we account the depth of the logical circuit, the number of qubits and whether or not to perform a measurement at the end of the circuit. Once a circuit is generated using this method, then it is transpiled. The transpile function allows us to define the universal set of gates which the circuit has to be transpiled to. As for the IBM machines, we employ the set [``rz", ``sx", ``x", ``cx"] from the Eagle backend, see the \href{https://quantum.cloud.ibm.com/docs/en/guides/get-qpu-information}{online guidelines}, while for the IonQ machines [``rxx", ``rz", ``ry", ``rx"]. An example of code used to perform the preparation of the circuit in terms of the native gates presented before, using qiskit libraries, is shown in Appendix \ref{app:snippet_example}. The explicit algebraic representation of the gates is provided in Appendix \ref{app:gates}.
The size of the training dataset is determined through empirical analysis of model convergence. Initially, we employ $3000$ samples per qubit count; however, the model struggles to capture both the syntactic grammar of OpenQASM and the underlying physical unitary properties. Increasing the dataset to 5,000 samples yielded a notable improvement in convergence, yet optimal performance remained out of reach. We ultimately determined that $10000$ samples per qubit were necessary to consistently achieve a valid circuit generation rate of approximately 95\%.

While the synthetic nature of the data allows for virtually unlimited dataset expansion, we identified $10000$ samples as the optimal trade-off between training computational cost and model performance. Consequently, our final training set comprises $10000$ circuit pairs for each qubit count. The specific method used to convert these raw circuits into machine-readable inputs, a process known as tokenization, is detailed in the following Section.

\subsubsection{\label{tokenizer}Tuning the tokenizer for the QASM standard}

Although often treated as a secondary component in architecture design, the tokenizer is the key building block of this work. Indeed, deep learning models cannot process raw text, since they require numerical inputs. The tokenizer bridges this gap by breaking text down into atomic units, called tokens, and converting them into numerical representations which the model can encode and learn from.
In standard Natural Language Processing (NLP), a balance must be struck~\cite{vaswani2017attention}. Assigning a token to every distinct word creates an unmanageably large vocabulary, while assigning a token to every letter strips away any semantic meaning. Therefore, NLP usually finds an optimal middle ground using sub-word units.

However, our application deals with Quantum Assembly (QASM), a formal programming language governed by a stricter syntax compared to natural speech patterns. Consequently, we diverge from the standard NLP methods. To adhere to the rigid grammar of QASM, we built our tokenizer using Regular Expressions (RegEx). In a few words, RegEx is a tool that defines specific search patterns, allowing us to systematically identify and extract meaningful grammatical structures from the code.
We therefore define specific patterns to capture essential components of QASM, as it can be seen in Appendix \ref{app:snippet_example}.
This RegEx-based approach allows for rapid parsing of QASM strings, transforming the raw code into a simplified meta-code. 
This intermediate representation removes superfluous syntactic noise -- for example, reducing the verbose ``q[1];'' to the atomic ``q1;''. 
By streamlining the grammar, we enable the model to focus on the underlying quantum features rather than rigid syntactic structures. 
However, a challenge remains when dealing with the angles of rotation of parametric gates $\hat{U}(\theta)$. 
While rotation parameters are inherently continuous, Transformer models operate on discrete entities i.e. tokens.
To address this issue, we introduce a discretization strategy that converts continuous angles into symbolic tokens vocabularies (symbols, i.e. tokens). 
We begin by reducing precision, rounding each angle to two decimal places so that values such as $2.1232$ become $2.12$.
The angles are then normalized through a modulo-$2\pi$ operation, ensuring that all values - whether initially positive or negative - are mapped into the interval $[0, 2\pi)$.
Once normalized, each angle is discretized by scaling it according to a grid size $\lambda=128$ and assigning it to an integer bin. This is done using
\begin{equation}
    i = \left\lfloor \frac{\theta_{norm}}{2 \pi}\cdot \lambda \right\rfloor
\end{equation}
where the floor function $\lfloor \cdot \rfloor$ ensures that the unit circle is divided into $\lambda$ uniform sectors, each corresponding to a unique symbolic token.
The effects of this process are shown in Appendix \ref{app:snippet_example}, where continuous angle rotations, such as ``rz(3.19)'', are mapped through the meta-code into a strict triple, i.e.: \texttt{<PARAM\_START> PARAM\_64 <PARAM\_END>}. Here, the framing tokens denote the start and end of a parameter sequence, while the middle token, \texttt{PARAM\_64}, represents the specific bin index $i=64$. 

By inverting the discretization formula, we can determine the angle associated with this token. In this particular case, the \texttt{PARAM\_64} token represents a rotation of:
\begin{equation}
    \theta_{rec} = \frac{64}{128} \cdot 2\pi = \pi \approx 3.14159
\end{equation}
Note that the original angle $3.19$ is approximated to $\pi$ due to the finite resolution $\lambda$ of the discretization grid.

\subsubsection{QASM Embedding}

After the tokenization step, which converts the QASM code into sequences of integer indices, the data is ready to be fed into the model. The initial stage of the architecture -- and a key factor in the success of Transformers in NLP -- is the projection of these discrete tokens into a learnable latent space.
Such latent space is the internal geometric representation of the model in which raw inputs (sequences of text for Transformers or images for convolutional neural networks \cite{li2021survey}, graphs or nodes for graph neural networks \cite{zhou2020graph, lee2019mathematical} and so on) are mapped to vectors where each component is a real value.
These continuous vectors capture the essential structure and meaning of the inputs by simply assigning a position in a huge dimensional space.
In fact, similar entities are mapped and located close to each other, allowing the model to reason about relationships, patterns, and context through Euclidean geometry, i.e., distances and directions, as shown in the second step of Figure \ref{fig:overallView}.
During training, backpropagation refines this mapping to produce token embeddings.
These dense vectors serve as the foundational input fed into the Transformer layers.

\subsubsection{Model beyond the transpiling transformer}

Transformers are well-suited for text processing because they use attention mechanisms to capture dependencies between words. While self-attention allows the model to estimate the next word based on preceding words in the same sequence, the mechanism suitable for our purposes is cross-attention. It is a modification of self-attention for which the model attends to the original source sequence (Encoder) while generating the new target sequence (Decoder). It allows the model to align input and output contexts, making cross-attention essential for tasks such as natural language translation. Essentially such process translates directly to the task of quantum transpilation, the model will attend to the input QASM while generating the transpiled QASM. It is a conceptual difference between standard decoder-only architecture.
A critical hyperparameter for this architecture is the \textit{context window} size, set here to 768 tokens. This feature defines the maximum sequence length for both the input QASM (Encoder) and the generated output (Decoder). If the tokenized representation of a circuit exceeds this window, the QASM code would be truncated, rendering the resulting circuit invalid. Given the strict syntactic requirements of QASM, we exclude these circuits from the training set entirely. This mechanism ensures the model is not exposed to truncated, malformed sequences, which could otherwise degrade its ability to generate syntactically valid code.
\begin{table}[H]
    \centering
    \begin{tabular}{lr}
        \hline \hline
        \textbf{Hyperparameter / Component} & \textbf{Value / Count} \\
        \hline
        \multicolumn{2}{c}{\textit{Global Configuration}} \\
        Embedding Dimension ($d_{\text{model}}$) & 768 \\
        Feed-Forward Dimension ($d_{\text{ff}}$) & 2,048 \\
        Attention Heads ($h$) & 8 \\
        Head Dimension ($d_k$) & 96 \\
        Context Window & 768 \\
        \hline
        \multicolumn{2}{c}{\textit{Attention Mechanism}} \\
        Attention Heads per Layer ($h$) & 8 \\
        Head Dimension ($d_k$) & 96 \\
        Total Attention Heads (Encoder) & 48 \\
        Total Attention Heads (Decoder) & 48 \\
        \hline
        \multicolumn{2}{c}{\textit{Encoder Stack (6 Layers)}} \\
        Self-Attention Parameters (per layer) & 2,362,368 \\
        Feed-Forward Network (per layer) & 3,150,080 \\
        Layer Normalization (per layer) & 1,536 \\
        \textbf{Total per Encoder Layer} & \textbf{5,513,984} \\
        \hline
        \multicolumn{2}{c}{\textit{Decoder Stack (6 Layers)}} \\
        Self-Attention Parameters (per layer) & 2,362,368 \\
        Cross-Attention Parameters (per layer) & 2,362,368 \\
        Feed-Forward Network (per layer) & 3,150,080 \\
        Layer Normalization (per layer) & 3,072 \\
        \textbf{Total per Decoder Layer} & \textbf{7,877,888} \\
        \hline
        \multicolumn{2}{c}{\textit{Total Model Size}} \\
        Trainable Parameters & 80,358,915 \\
        Non-Trainable Parameters & 0 \\
        \textbf{Total Parameters} & \textbf{80.36 M} \\
        \hline \hline
    \end{tabular}
    \caption{Hyperparameters and architectural details of the Transformer model used for transpilation.}
    \label{tab:model_parameters}
\end{table}
Further details about the transformers are provided in Appendix \ref{sec:appendix_transformer}.

\subsection{\label{sec:loss}Training loss and accuracy setting}

The model is trained by optimizing a composite loss function, combining the standard token-based Cross-Entropy (CE) loss ($\mathcal{L}_{CE}$) with a physics-informed fidelity loss ($\mathcal{L}_F$), weighted with $\alpha$ and $\beta$, respectively. Moreover, in Section \ref{sec:Metrics}, we introduce some metrics which, even though not employed to train the model and update its parameters, they are still correlated to the loss terms and worthy of considerations when assessing the goodness of the results.

\paragraph{Token-based Cross-Entropy Loss}

The primary loss function is the standard cross-entropy loss, $\mathcal{L}_{CE}$, typical for token-based sequence generation. At each step, the model outputs a probability distribution $\hat{\mathbf{y}}$ over the entire vocabulary of $V$ possible next tokens.
Given the true one-hot target vector $\mathbf{y}$, the loss for a single token prediction simplifies to the negative log-probability of the single correct token $c$:
\begin{equation}
    - \log(\hat{y}_c)
\end{equation}
The above loss penalizes the model for assigning low probability to the correct token. The total loss for a circuit is the average of $\mathcal{L}_{CE}$ over the entire sequence, i.e.
\begin{equation}
\label{eq:cross_entropyLoss}
    \mathcal{L}_{CE} = - \frac{1}{N} \sum_{i=1}^N \log(\hat{y}_i)
\end{equation}

\paragraph{Fidelity Loss}

Let \( U_{\text{ref}} \) and \( U_{\text{pred}} \) be the reference and predicted unitary matrices, respectively.
The fidelity F between \( U_{\text{ref}} \) and \( U_{\text{pred}} \) is calculated as follows:
\begin{equation}
\label{eq:fidelity}
    F(U_{\text{ref}}, U_{\text{pred}}) = \frac{1}{d^2} \left| \text{Tr}(U_{\text{ref}}^\dagger U_{\text{pred}}) \right|^2
\end{equation}
Consequently, the fidelity loss $\mathcal{L}_F$ we aim to minimize is:
\begin{equation}
\label{eq:lossFidelity}
    \mathcal{L}_F = 1 - F = 1 - \frac{1}{d^2} \left| \text{Tr}(U_{\text{ref}}^\dagger U_{\text{pred}}) \right|^2
\end{equation}

\paragraph{Loss combination}

\begin{figure}[htbp]
    \centering
    \includegraphics[width=0.5\textwidth]{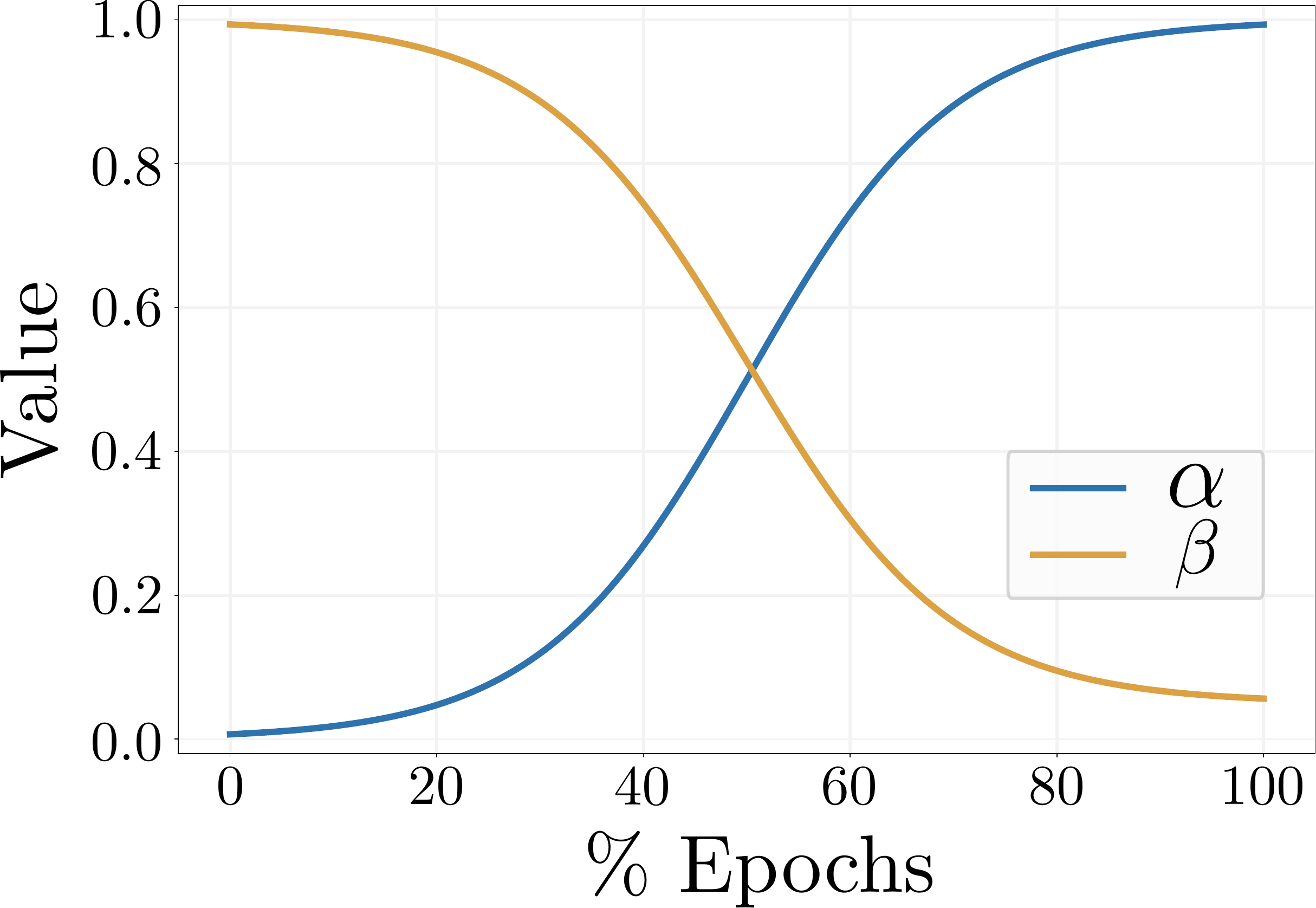}
    \caption{Scheduling with dynamical loss combination, which prioritize the grammar aspect of the learning first and thereafer the physical meaning of the produced QASM. Due to the label smoothing techniques, the cross-entropy loss $\mathcal L_{CE}$ never decreases to zero, but converges to a small value $\epsilon>0$.}
    \label{fig:loss_parameters}
\end{figure}

In our tests, it does not emerge one predominant configuration to the others. This fact can be ascribed to the rounding of the rotations. For this simpler rotation angle it is sufficient to have the token loss which over time will indirectly learn also the physical interpretation. It remains to be seen whether or not with less rounding, larger model the fidelity may help more the model converge. In Figure \ref{fig:loss_parameters} the combination of losses used in the experiments are shown . It follows the expression for the linear combination of the loss functions:
\begin{equation}
\label{eq:totalLoss}
    \mathcal{L} = \alpha\,\mathcal{L}_F + \beta\,\mathcal{L}_{CE}
\end{equation}

\paragraph{Label smoothing}

To prevent the model from becoming overconfident and to encourage better generalization, we apply \textit{label smoothing} \cite{muller2019does} to the cross-entropy objective. 
Instead of training on hard one-hot targets (where the correct token probability is exactly 1.0), we smooth the target distribution using a factor $\epsilon$. 
The new target probability $y^{LS}_c$ for the correct class becomes $1 - \epsilon$, while the remaining probability mass $\epsilon$ is distributed uniformly across the vocabulary:
\begin{equation}
    y^{LS}_i = \begin{cases} 
    1 - \epsilon & \text{if } i = c \\
    \frac{\epsilon}{V-1} & \text{if } i \neq c
    \end{cases}
\end{equation}
Here $V$ represents the size of the vocabulary, while $c$ represents the index in the vocabulary that corresponds to the correct prediction.
This regularization fundamentally alters the convergence limit. 
Unlike standard cross-entropy, where a perfect model drives the loss to zero, a model trained with label smoothing converges to a non-zero floor. 
This behavior is shown in Figure \ref{fig:loss_parameters}, where the cross-entropy loss does not converge asymptotically to zero, but rather to an upper value of $\epsilon>0$.
Because the target distribution itself contains uncertainty (non-zero entropy), even a model that predicts the targets perfectly will retain a loss value equal to that inherent entropy.

\paragraph{Accuracy Metric}

In our experimental analysis, we utilize this behavior to define the accuracy of the model. Specifically, the lower the total loss $\mathcal{L}$, the higher the accuracy when training the model, even when we introduce a theoretical non-zero lower bound on $\mathcal L$ due to label smoothing techniques. Unlike a simple token-matching percentage, the metric $\mathcal{L}$ in Equation \eqref{eq:totalLoss} provides a more holistic view of the training progress, as it encapsulates both the syntactic correctness (via $\mathcal{L}_{CE}$) and the physical correctness (via $\mathcal{L}_F$). Therefore, in this work, the \textit{converged accuracy} corresponds to the combined loss stabilizing at its minimum smoothed value, and it is employed to train the model.

\subsubsection{\label{sec:Metrics}Metrics to assess the correctness of the training}

Let's now introduce a set of metrics employed during the training to evaluate the goodness of the results. Still, the metrics are not exploited by the model to perform backpropagation, but rather to assess whether the training has been successful or not.

\paragraph{Grammar accuracy}

We introduce the feature of \textit{grammar accuracy}, i.e. a score in the range $\{0,1\}$ to assess the correctness of the grammar of the output QASM files. In other words, if the IBM compiler is able to convert back the output QASM file from the transformer to a quantum circuit, the grammar accuracy is equal to $1$, otherwise $0$. Therefore, it is straightforward to introduce this metric as the percentage of correctly translated QASM files under the grammar point of view. Take notice, this metric does not check the fidelity of the unitary corresponding to the QASM output file.

\paragraph{Perplexity}

When dealing with NLP, the perplexity is a known metric defined as the exponential expression for the cross-entropy $\mathcal{L}_{CE}$ defined in Equation \eqref{eq:cross_entropyLoss}~\cite[pg.38]{jurafsky2020speech}:
\begin{equation}
    \text{Perplexity} = \exp(\mathcal{L}_{CE}) = \exp\left(- \frac{1}{N} \sum_{i=1}^N \log(\hat{y}_i)\right)
\end{equation}
Such metric quantifies the statistical surprisal \cite{modirshanechi2022taxonomy} of the model regarding the target data. Intuitively, a perplexity value of $k$ indicates that the model is, on average, as uncertain as if it were choosing uniformly among $k$ candidates for the next token. A perfect training would entail $\hat y_i = 1 \, \forall i$, therefore the perplexity would be equal to $1$.
Consequently, a lower perplexity signifies that the model is assigning higher probability mass to the correct tokens. The goal during training is to drive this value as close as possible to the theoretical minimum of 1, which would represent absolute certainty and perfect prediction.

\subsection{Benchmarking parametric continuous circuits versus a universal discrete set of gates}\label{Solovay} 

In this Section, we describe a complementary approach for transpiling via transformers, aimed to map a circuit decomposed through the Solovay-Kitaev algorithm from the IBM native gates to the IonQ ones. Here, the IBM parametric circuit, depending on rotation angles such as $\pi/2$, $\pi/4$, is mapped into a circuit composed by a discrete but universal set of native gates, as summarized in Figure \ref{fig:overallView}.
Indeed, in the case discussed in Section \ref{tokenizer}, the circuit is composed by Clifford gates (e.g. Hadamard gates and CNOT operators) plus a set of rotations (involving a single qubit or two qubits). In the case we are dealing with in this Section, we morph the original circuit into its Solovay-Kitaev version, i.e. all the single and two-qubits rotations are decomposed in a universal set of gates which do not require any continuous parameter such as rotation angles. An illustrative example for such an universal set is $\{\hat H, \hat T, \hat T^\dagger \}$. 


The meaning and the notation used to express the Solovay-Kitaev theorem is reported in Appendix \ref{app:SKscalingLaws}. Here, we inquire the scaling laws of complexity of the transformer model when translating a parametric circuit to another set of (continuous and parametric) gates, and its Solovay-Kitaev decomposed version into the corresponding one. From the aforementioned Appendix, we prove that the complexity law for the tokenization, when approaching the transpiling process on a Solovay-Kitaev decomposed circuit, is expressed by Equation \eqref{eq:SKTransformerscaling}.
If the number of gates scales as $f(n)$, $n$ being the number of qubits in the register, and $\varepsilon$ the precision we want to achieve through the Solovay-Kitaev decomposition, we can state that the complexity of the model, once the quantum circuit has been decomposed though the Solovay-Kitaev algorithm, would scale as
\begin{equation}
\label{eq:SKTransformerscaling}
    O\left(nf(n) \log^c\left(\frac{n f(n)}{\varepsilon}\right)\right)
\end{equation}

\paragraph{Continuous rotations versus Solovay-Kitaev: scaling laws}

A critical point of the proposed algorithm consists of inquiring its complexity laws, i.e. assessing how the size of the neural network scales compared to the size of the quantum circuit. In the first place, we must address the number of tokens to identify the action of a single gate. In more detail, any gate should be distinguished for the qubit it is acting on and which kind of gate we are employing.
Just for instance, suppose a Hadamard gate is acting over the $i$-th qubit $\hat H_i$, therefore, a proper tokenization should allow the net to distinguish $i$ and $H$ as input parameters. When dealing with the quantum circuit, two parameters thus arise, $N_g$ and $N_q$, i.e., respectively, the number of available gates and the number of qubits in the register. Take notice, $N_g$ should consider the gates of both the hardware platforms, $N_g = N_{{source}} + N_{{target}}$. In the list of available gates provided for IBM and Rigetti in Section \ref{app:gates}, we see that $N_{source} = 4$ and $N_{target} = 5$, thus $N_g = 9$.

A second criterion concerns parametrized rotation gates $\hat U(\theta)$, that is the rotations $\hat U$ with angles $\theta \in [0;2\pi]$. Our approach consists in sampling a number $2\pi/\epsilon_\vartheta$ of angles, tuning a precision $\epsilon_\vartheta$. Therefore, the final parameter $N_{ang}$ describes the number of tokens required to identify a specific angle, with accuracy $\varepsilon_\vartheta = 2\pi\epsilon_\vartheta$. We can thus consider the number of tokens required to sample the angles as $N_\vartheta = \lceil 2 \pi/\epsilon_\vartheta \rceil$, and the precision on the unitary as follows:
\begin{equation}
    \forall \varepsilon_\vartheta \; \exists \, \delta > 0: \left| \vartheta - \vartheta_{approx} \right| < \varepsilon_\vartheta, \;d(\hat U(\vartheta), \hat U(\vartheta_{approx})) < \delta
\end{equation}
with $d(\cdot, \cdot)$ being a proper distance in the space of the unitary matrices $SU(2)$, e.g. the trace distance or the diamond norm~\cite{nielsen2010quantum}. The total number of tokens, required to characterize a quantum logic gate, can be instead described as
\begin{equation}
\label{eq:TokenLength}
    L = N_q + N_g + P_{ang}
\end{equation}
%



\begin{figure}[htbp]
    \centering
    \includegraphics[width=\textwidth]{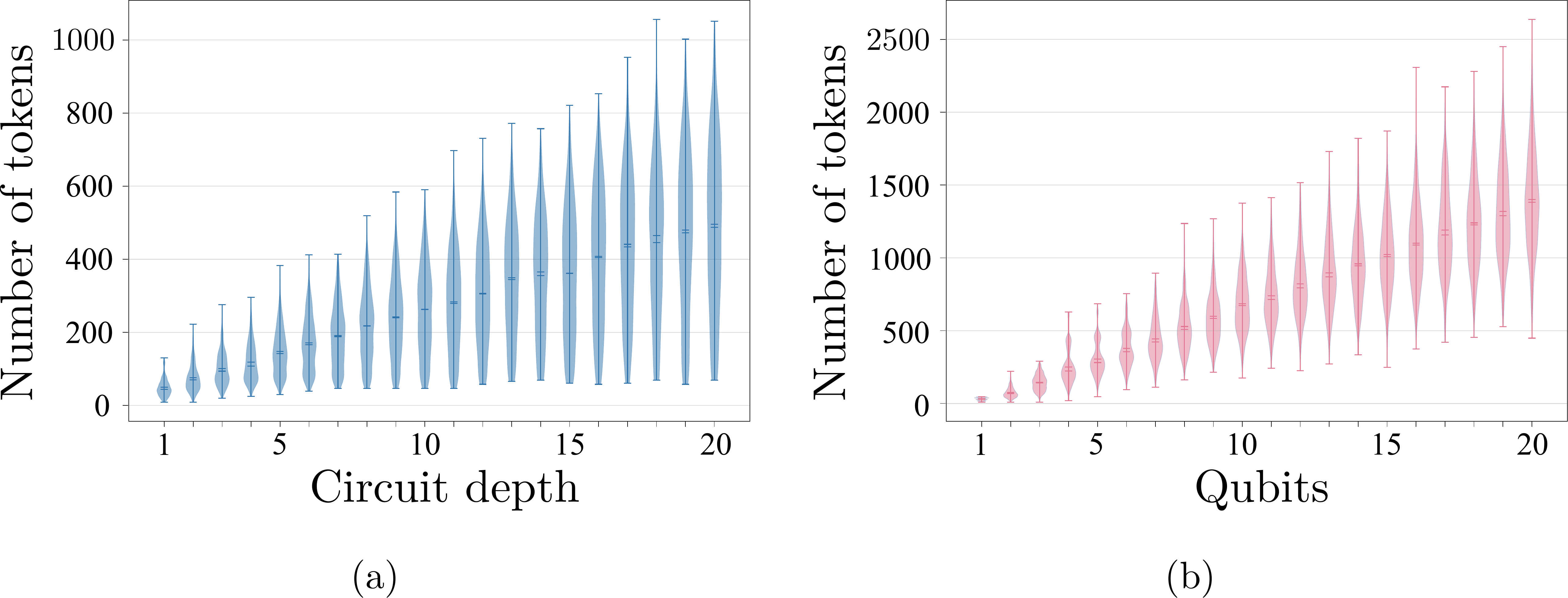}
    \caption{Count of the tokens as the depth of the circuit and as the number of qubits increase. In both cases, the corresponding quantity is kept fixed. In both cases, we observe a linear scaling, just as predicted in Equation \eqref{eq:circuitTokenLength}.}
    \label{fig:token_panel}
\end{figure}

Suppose now to deal with a circuit of length $m$, which means it can be described as a sequence of gates $\hat U_1 \circ \hat U_2 \circ ... \circ \hat U_m$. The overall size  $S$ of the transformer model can thus be described as
\begin{equation}
\label{eq:circuitTokenLength}
    S = m L = m (N_q + N_g + P_{ang})
\end{equation}
The numerical assessment of the above scaling law is reported in Figure \ref{fig:token_panel}, showing the trend when increasing both the number of qubits and the circuit length in the number of gates. To evaluate the number of tokens required, we first transpose the QASM file into its corresponding tokenized version, fixing time by time the maximum number of gates and the qubits in the register. Therefore, it is possible to empirically characterize the law in Equation \eqref{eq:circuitTokenLength} without training the model itself, but rather just preparing the encoding for the training.

\section{\label{sec:results}Experiments and results}

We now turn to the evaluation of the transformer-based transpiler across three distinct scenarios to test its generalization capabilities, namely Cross-Platform transpilation (IBM to IonQ), intra-platform optimization (IBM Eagle to Heron), and discrete gate decomposition (Solovay-Kitaev), respectively. 
For the Solovay-Kitaev approach, we transpile the standard native set of gates from IBM, $\{ \hat T, \hat T^\dagger, \hat H\}$, to the IonQ one $\{ \hat S, \hat S^\dagger, \hat H\}$, see Figure \ref{fig:overallView}. 
For all of these scenarios, the same model (i.e. the same architecture and the same number of parameters) and the same loss is used in order to get comparable results. These parameters are reported in Table \ref{tab:model_parameters}.
As for training the models, we employed as metrics the grammar accuracy of the model, the fidelity and the perplexity. We refer to Section \ref{sec:Metrics} for the definition of grammar accuracy and perplexity, while to Equation \eqref{eq:fidelity} for the fidelity. 
Take notice, while the fidelity loss $\mathcal L_F$ in Equation \eqref{eq:lossFidelity} is employed for training the model, the fidelity $F$ is a quantum feature to check whether the output circuit and the original one implement the same unitary operator.

\begin{figure}[H]
    \centering
    \includegraphics[width=\linewidth]{ 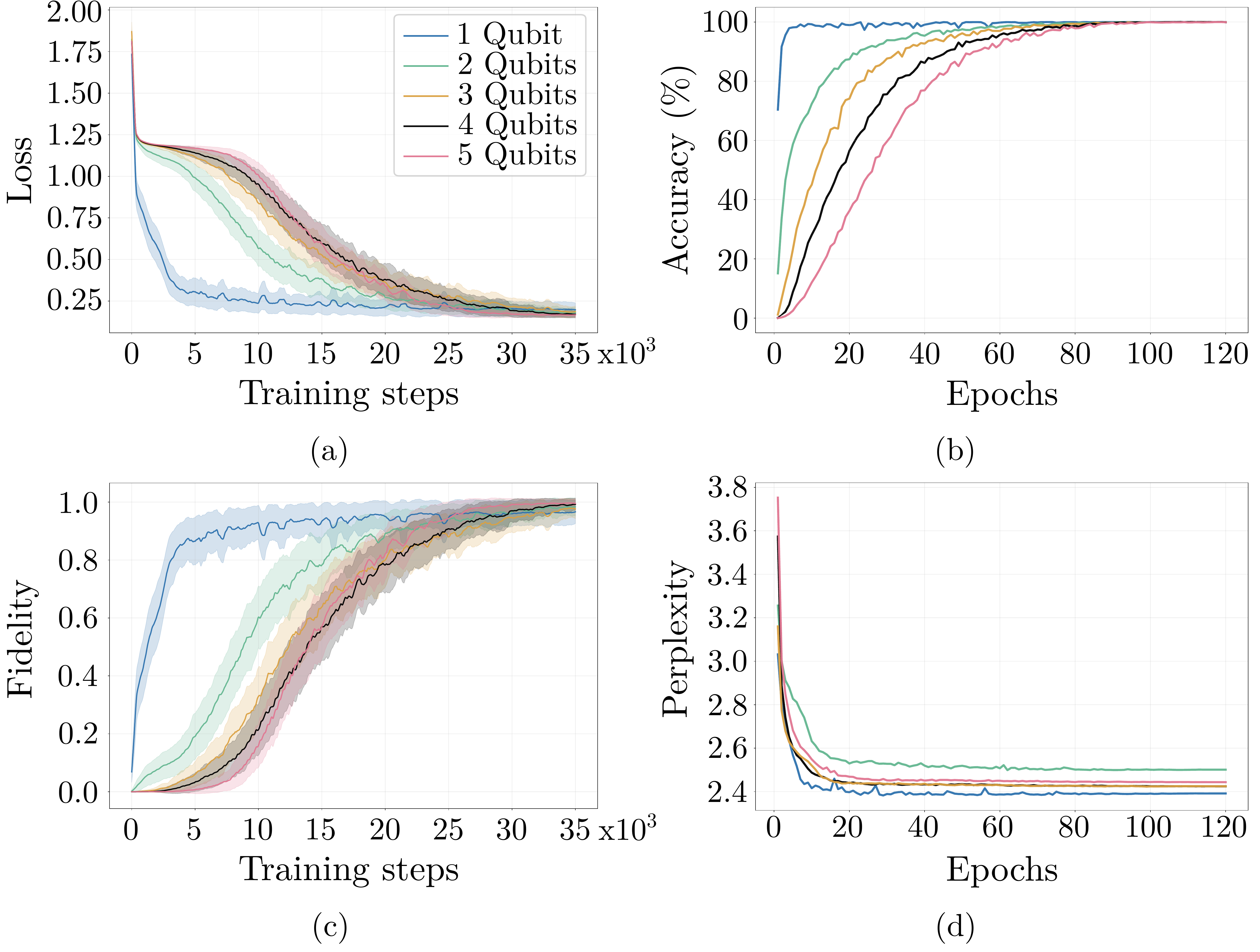}
    \caption{Training metrics overview. (a) Loss evolution throughout the training steps. (b) Grammar accuracy throughout the epochs of training. (c) Circuit fidelity evolution during the training steps. (d) Perplexity evolution during the training epochs.}
    \label{fig:training_results_combined}
\end{figure}

As shown in Figure \ref{fig:training_results_combined}, the first scenario (i.e. transpilation between two different quantum platforms) is successful across all qubit counts (1 through 5), by converging in terms of both fidelity between the unitaries (original and target) and accuracy. The loss in Figure \ref{fig:training_results_combined}(a) is shown to decrease asymptotically to zero, indicating that the training of the model is successful. However, even though always converging to zero, we can observe a different trend between the loss related to 1-qubit circuits and the losses related to the multiple-qubits circuits. This behavior can be explained by considering that, when involving a register of $n \geq 2$ qubits, the entangling gates enter the vocabulary of the transformers, which hinders the process of transpiling for the transformers. For this reason, a sort of discontinuity occurs when switching from one to multiple qubits in Figure \ref{fig:training_results_combined}.
To highlight the goodness of the training, we now show the metrics beyond the loss function itself.
We display the grammar accuracy, i.e. the number of QASM files correctly translated from a grammatical point of view, in Figure \ref{fig:training_results_combined}(b). 
Since the grammar of QASM files is correctly compiled almost $100\%$ of times, we can focus on whether the output unitaries match the original ones, i.e. the transpiled IonQ circuits correspond to their IBM counterparts. This flag can be retrieved by checking the fidelity, whose trend throughout the training is reported in Figure \ref{fig:training_results_combined}(c). Regarding the last metric, the perplexity, we can notice from Figure \ref{fig:training_results_combined}(d) that it fluctuates between $2.4$ and $2.6$. This outcome stresses a strong confidence for the model when predicting the next token.

\begin{figure}[H]
    \centering
    \includegraphics[width=\linewidth]{ 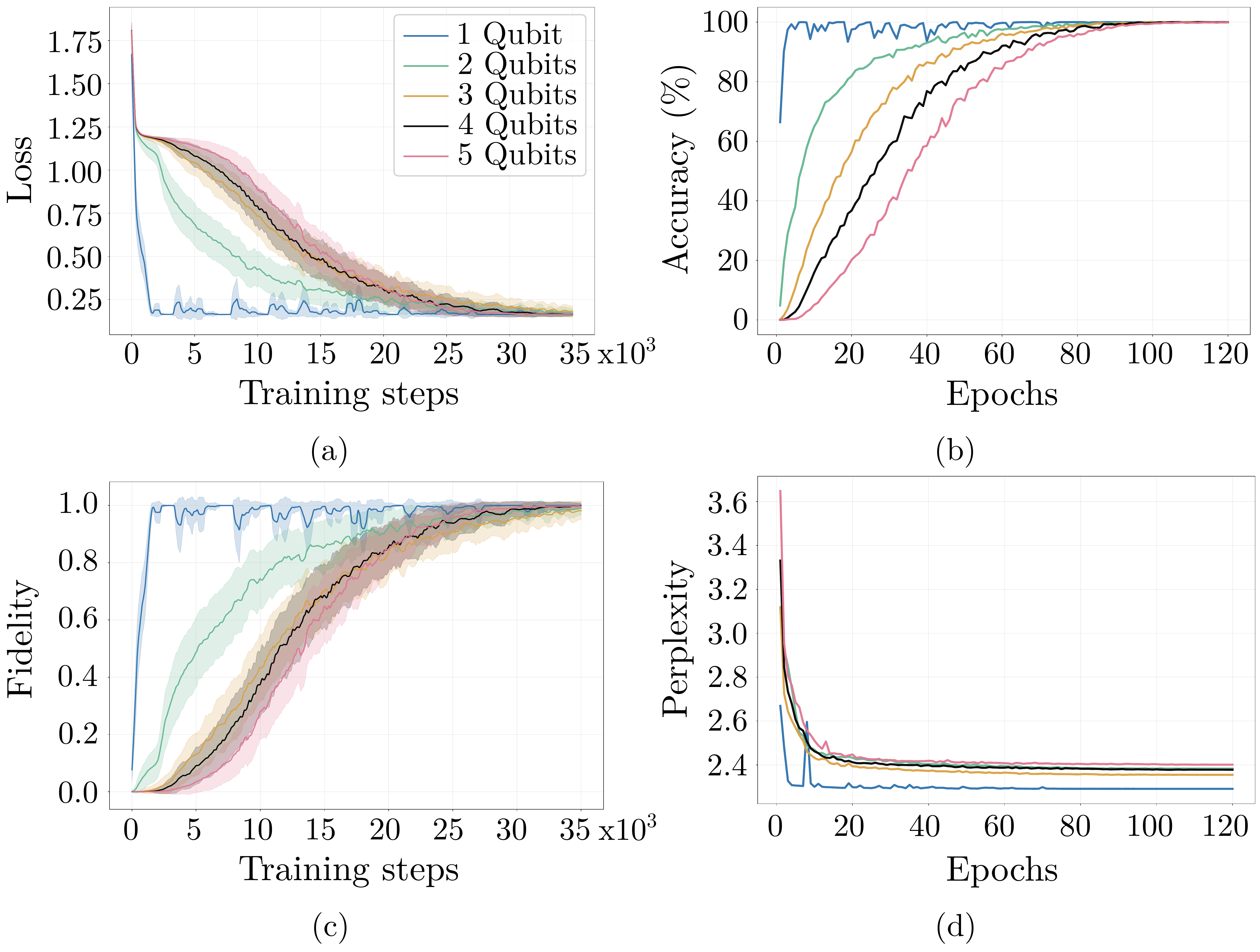}
    \caption{Training metrics overview when transpiling between IBM Eagle and IBM Heron backends. (a) Loss evolution thoughout the training steps. (b) Grammar accuracy throughout the epochs of training. (c) Circuit fidelity evolution during the training steps. (d) Perplexity evolution during the training epochs.}
    \label{fig:BAECKEND-training_results_combined}
\end{figure}

The results of the first scenario (IBM Eagle to IonQ) are also repeated in the second scenario (IBM Eagle to IBM Heron), shown in Figure \ref{fig:BAECKEND-training_results_combined}. 
Here, we see  the same behavior showcased in the first scenario, highlighting the ability of this model to learn also subtle changes when transpiling from an IBM backend (Eagle) to a very similar one (Heron).

\begin{figure}[H]
    \includegraphics[width=\linewidth]{ 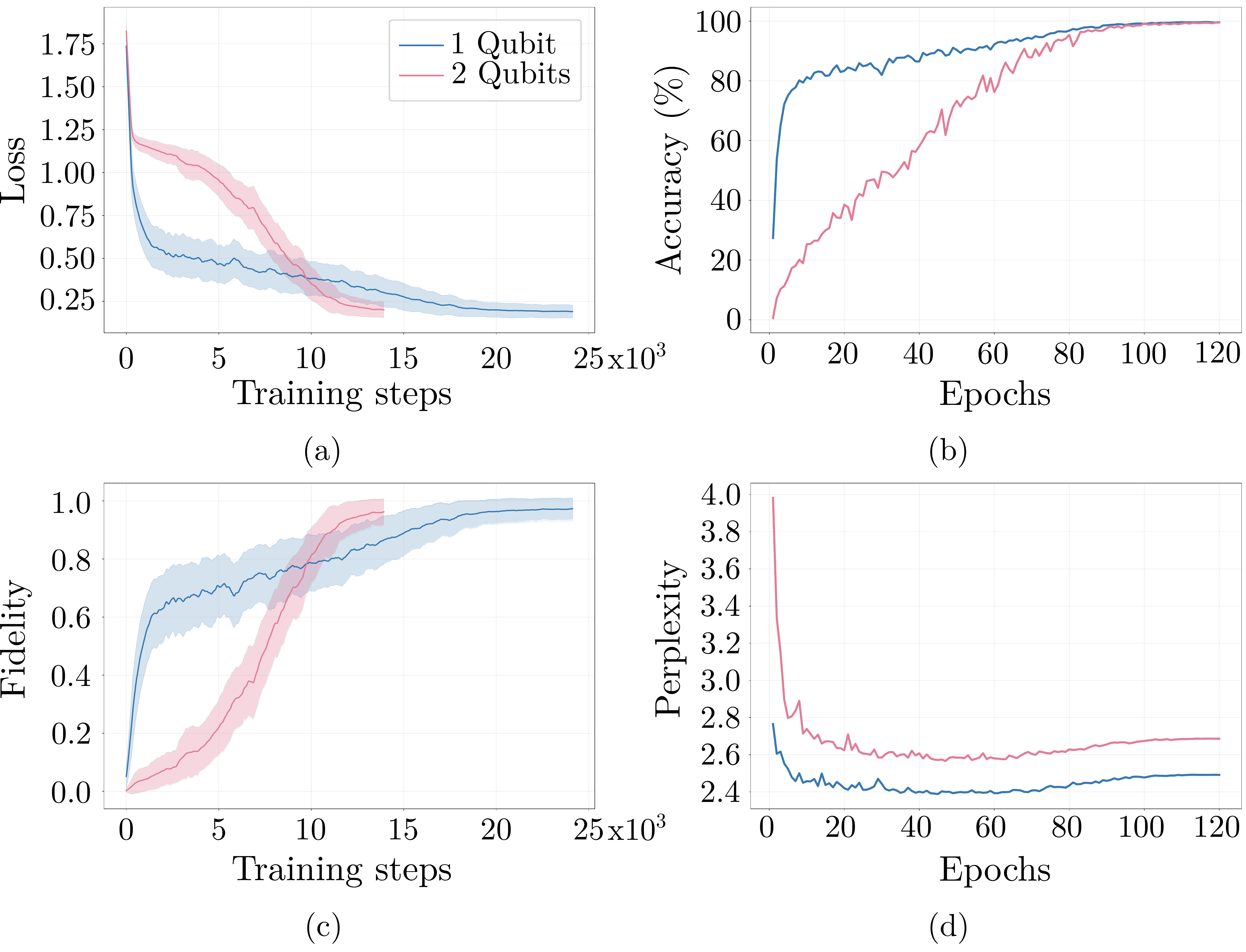}
    \caption{Training metrics overview for training the transpiler transformers on the Solovay-Kitaev circuits. The register of these circuits is up to $2$ qubits. Since the training set for the $2$-qubits circuits is smaller than the $1$-qubit ones, the $2$-qubit training ends 10000 steps before their $1$-qubit counterparts. (a) Loss evolution. (b) Grammar accuracy. (c) Circuit fidelity. (d) Perplexity.}
    \label{fig:SOLOVAY-training_results_combined}
\end{figure}

Finally, in the third scenario, a computational bottleneck was hit. Due to the characteristics of the Solovay-Kitaev decomposition, a portion of the training dataset exceeds the model's fixed context window of 768 tokens (see Table \ref{tab:model_parameters}), consequently, successful training was limited to 1 and 2 qubit circuits. 
These results highlight a constraint in the model dimension and the number of parameter required. 
When applying the Solovay-Kitaev decomposition, the sequence length $m$ scales according to $O(m \log^c(m/\epsilon))$, see Equations \eqref{eq:SKscaling}, \eqref{eq:SKTransformerscaling}. 
Unlike the standard IBM-to-IonQ mapping, this decomposition induces a rapid expansion in token count. 
Consequently, during the dataset generation for 3, 4, and 5-qubit registers, a statistically significant majority of the decomposed circuits exceeded the fixed context window.
We deliberately choose not to filter for shorter outlier circuits to artificially populate the dataset, as this would introduce selection bias and result in a model trained on non-representative, low-depth decompositions. 
Instead, we report results for 1 and 2 qubits, where most of the sequence lengths remain compatible with the architecture. This behavior emerges from Figures \ref{fig:SOLOVAY-training_results_combined}(a) and \ref{fig:SOLOVAY-training_results_combined}(c).
This limitation is not a failure in the learning capability of the transformers, but rather an experimental confirmation of the steep polynomial scaling of token requirements for discrete basis compilation. 
This trend highlights that, while standard transformer models can handle continuous parameterized gates efficiently, discrete decomposition requires significantly scaled infrastructure (HPC) and extended context windows.
Nevertheless, we remind that the precision when tokenizing the angles is fixed to $10^{-2}$, see Section \ref{tokenizer}, but could be further increased following the scaling law from Equation \eqref{eq:circuitTokenLength}. 
Finally, the empirical results from Figure \ref{fig:SOLOVAY-training_results_combined} confirm the theoretical scaling laws proposed in Section \ref{Solovay}. 
The token usage scales polynomially with circuit depth and qubit count.
However, the Solovay-Kitaev experiment demonstrates that, when the gate decomposition increases in its sequence length, the overhead shifts from the learnability of the model to the memory boundaries, when storing and reprocessing the QASM files.

\section{\label{sec:conclusions}Conclusions}

We demonstrate that transformer-based architectures can effectively perform quantum circuit transpiling, handling the problem as a sequence-to-sequence translation task. The model achieves success rates exceeding 99.98\% for continuous rotation gates up to 5 qubits. 
While the model excels at continuous parameter mapping, our experiments with Solovay-Kitaev decomposition reveal the limitations of standard context windows. The increasing of the sequence lengths necessitates models with significantly larger attention spans, which in turn requires scaling to classical larger hardware resources. Nevertheless, the resolution on the angles of the rotations is fixed up to $10^{-2}$. By increasing with the precision, the scaling laws in Equations \eqref{eq:TokenLength} and \eqref{eq:circuitTokenLength} could be assessed by a numerical benchmark.
To conclude, transformer-based approach proved promising and opens new research and application paths for automated and systematic transpiling tasks.   

\section*{Acknowledgments}

We owe thanks to Lorenzo Moro for the fruitful discussions about transformers in the field of quantum compiling. 
The authors acknowledge support from the Qgraph project funded via the National Centre for HPC, Big Data and Quantum Computing
(HPC) (grant No. D43C22001240001 CN00000013). P. Z. and E. P. are grateful to NTT Data for having co-funded the D.M. 117 PNRR PhD grant.

\bibliographystyle{apsrev4-2}
\bibliography{bibliography}

\appendix
\section{The Transformer Model}
\label{sec:appendix_transformer}

The model used in this work is a \textbf{transformer}, an architecture that has become the state-of-the-art for sequence-to-sequence (seq2seq) tasks, such as machine translation. Our task, transpiling from one QASM standard to another, is a perfect example of such a translation task.

\begin{figure}[H]
    \centering
    \includegraphics[scale=0.7]{ 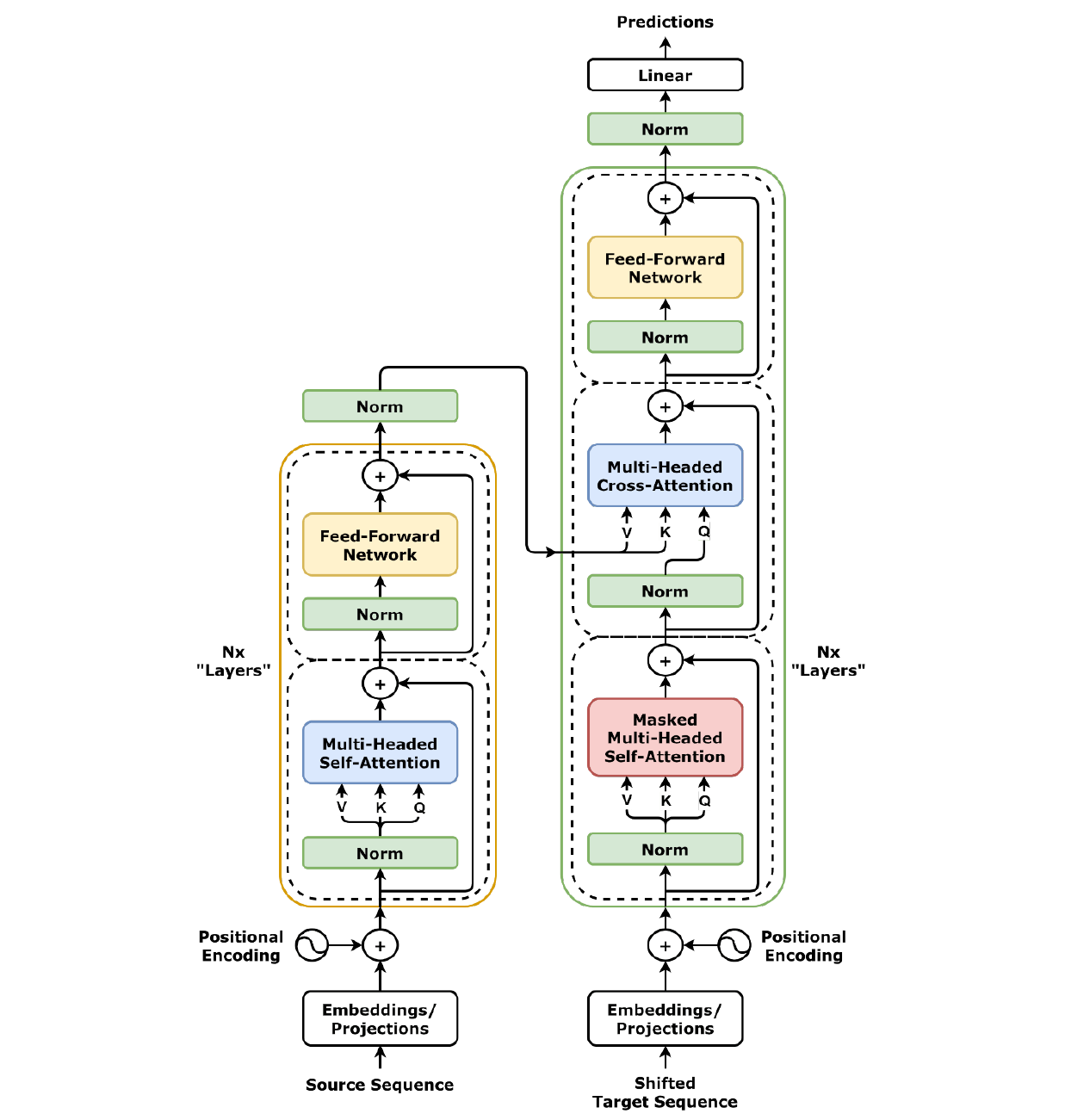}
    \caption{Encoder-Decoder architecture. Image by \href{https://github.com/dvgodoy/dl-visuals}{dvgodoy} / \href{https://creativecommons.org/licenses/by/4.0/}{CC BY}}
    \label{fig:transformerFull}
\end{figure}

The specific architecture is an \textbf{encoder-decoder} model, as depicted in Figure \ref{fig:transformerFull}

This design consists of two primary components:
\begin{itemize}
    \item \textbf{The Encoder (Left Side):} Its job is to read, understand, and process the \textit{entire} input sequence (the source IBM QASM circuit). It learns all the relationships and dependencies between the gates and qubits in the input and compresses this understanding into a set of context-rich vectors.
    \item \textbf{The Decoder (Right Side):} Its job is to generate the new, \textit{target} sequence (the destination IonQ QASM circuit), one token at a time. It does this by looking at what it has already written (its own output) and, crucially, by consulting the context-vectors passed to it by the encoder.
\end{itemize}

This entire process is built upon a few key concepts: vector embeddings, positional encoding, and the attention mechanism.

\subsection*{Input Processing: From QASM to Vectors}

A neural network cannot process raw text like \texttt{`rz(1.57) q[0];`}. Its inputs must be numerical. The first two steps convert our circuit-text into a format the model can understand.

\begin{itemize}
    \item \textbf{1. Tokenization:} We first break the raw QASM string into a sequence of discrete tokens. A described in Section \ref{tokenizer}, these tokens represent the fundamental units of our language, such as \texttt{`rz`}, \texttt{`q[0]`}, or a token representing a discretized angle.

    \item \textbf{2. Token Embedding:} We create a large lookup table, or embedding matrix, where each unique token in our vocabulary is mapped to a high-dimensional vector (e.g., 512 dimensions). This vector is \textit{learned} during training. This step turns our sequence of tokens (e.g., `[27, 8, 102]`) into a sequence of rich, dense vectors.

    \item \textbf{3. Positional Encoding:} A critical feature of transformers is that they process all tokens in a sequence simultaneously, unlike older models (like RNNs) that process them one by one. This parallel processing is highly efficient but loses all information about *word order*. For a quantum circuit, the order of gates is essential.
    
    To solve this, we inject a positional encoding vector. This is a fixed (not learned) vector that mathematically represents a token's position in the sequence (e.g., 1st, 2nd, 3rd...). This position vector is simply added to the token's embedding vector. The result is a final input vector that contains information about both \textbf{what} the token is (its semantic meaning) and \textbf{where} it is in the sequence (its position). The overall picture can be seen in the Example \ref{eq:inputModel}
\end{itemize}

\begin{equation*}
\begin{array}{r c c c}
    \textbf{1. Tokenization} & \texttt{"rz"} & \texttt{"1.57"} & \texttt{"q[0]"} \\
    \textit{(Map to Index)} & \downarrow & \downarrow & \downarrow \\
    \text{Token IDs} & [27] & [8] & [102] \\[1em]
    
    \textbf{2. Embedding} & \downarrow & \downarrow & \downarrow \\
    \textit{(Lookup Vector)} & 
    \begin{bmatrix} 0.1 \\ -0.5 \\ \vdots \end{bmatrix}_{\mathbf{e}_{27}} & 
    \begin{bmatrix} 0.9 \\ 0.2 \\ \vdots \end{bmatrix}_{\mathbf{e}_{8}} & 
    \begin{bmatrix} -0.3 \\ 0.4 \\ \vdots \end{bmatrix}_{\mathbf{e}_{102}} \\[2em]
    
     & + & + & + \\
    \textbf{3. Positional Enc.} & 
    \begin{bmatrix} \sin(p_1) \\ \cos(p_1) \\ \vdots \end{bmatrix}_{\mathbf{p}_1} & 
    \begin{bmatrix} \sin(p_2) \\ \cos(p_2) \\ \vdots \end{bmatrix}_{\mathbf{p}_2} & 
    \begin{bmatrix} \sin(p_3) \\ \cos(p_3) \\ \vdots \end{bmatrix}_{\mathbf{p}_3} \\[2em]
    \hline \\[-0.8em]
    
    \textbf{Model Input} & 
    = \mathbf{h}_1 & 
    = \mathbf{h}_2 & 
    = \mathbf{h}_3
\end{array}
\label{eq:inputModel}
\end{equation*}

\subsection*{The Encoder Stack}

The encoder's goal is to convert the sequence of input vectors $\mathbf{x} = (x_1, \dots, x_N)$ into a sequence of context vectors $\mathbf{z} = (z_1, \dots, z_N)$. Each vector $z_i$ in the output sequence has understood all the other vectors in the input. The encoder is a stack of $N_x$ identical layers, and each layer contains two sub-layers.

\subsubsection*{Sub-layer 1: Multi-Head Self-Attention (MHSA)}
This is the most important component of the transformer. Self-Attention is a mechanism that allows every token in the input sequence to look at and weigh the importance of every other token in that same sequence.

For each token's vector, we create three new vectors by multiplying it with three learned weight matrices:
\begin{itemize}
    \item \textbf{Query ($Q$):} "What information am I (this token) looking for?"
    \item \textbf{Key ($K$):} "What information do I (this other token) have?"
    \item \textbf{Value ($V$):} "If you (the Query) match me (the Key), here is the actual information (the Value) I will provide."
\end{itemize}

The model then calculates a score for every token against every other token by taking the dot product of their $Q$ and $K$ vectors ($Q \cdot K^T$). This score represents how relevant token $j$ is to token $i$. These scores are scaled (divided by $\sqrt{d_k}$) and passed through a $\text{softmax}$ function to turn them into weights that sum to 1.

Finally, the new vector for token $i$ is calculated as the weighted sum of all the $V$ vectors in the sequence. The result is that the new vector for \texttt{`q[0]`} has absorbed information from all the gates that act on it, and the vector for a \texttt{`cx`} gate has absorbed information from the \texttt{`q[0]`} and \texttt{`q[1]`} tokens it controls.

This is formally calculated as:
$$
\text{Attention}(Q, K, V) = \text{softmax}\left(\frac{QK^T}{\sqrt{d_k}}\right)V
$$
This process is multi-head because we do it $h$ times in parallel, each with different $Q, K, V$ weight matrices. This allows the model to learn different \textbf{types} of relationships simultaneously (e.g., one head might track gate-qubit relationships, another might track gate-angle relationships).

\begin{align*}
    \intertext{\textbf{Step 1: Calculate Relevance (The $Q \cdot K^T$ Matrix Mult.)}}
    \underbrace{
    \begin{bmatrix}
        \rule[2pt]{10pt}{0.5pt} \mathbf{q}_{\texttt{rz}} \rule[2pt]{10pt}{0.5pt} \\
        \rule[2pt]{10pt}{0.5pt} \mathbf{q}_{\texttt{q[0]}} \rule[2pt]{10pt}{0.5pt} \\
        \vdots
    \end{bmatrix}}_{Q \text{ (What I look for)}}
    \times
    \underbrace{
    \begin{bmatrix}
        | & | & \\
        \mathbf{k}_{\texttt{rz}} & \mathbf{k}_{\texttt{q[0]}} & \dots \\
        | & | & 
    \end{bmatrix}}_{K^T \text{ (What others offer)}}
    &= 
    \underbrace{
    \begin{bmatrix}
        \text{High} & \text{Low} & \dots \\
        \text{Low} & \text{High} & \dots \\
        \vdots & \vdots & \ddots
    \end{bmatrix}}_{\text{Relevance Scores}} \\[2em]
    \intertext{\textbf{Step 2: Aggregate Information (Weighted Sum of $V$)}}
    \underbrace{
    \begin{bmatrix}
        0.85 & 0.15 & \dots \\
        0.10 & 0.90 & \dots \\
        \vdots & \vdots & \ddots
    \end{bmatrix}}_{\text{Softmax(Scores)}}
    \times
    \underbrace{
    \begin{bmatrix}
        \rule[2pt]{10pt}{0.5pt} \mathbf{v}_{\texttt{rz}} \rule[2pt]{10pt}{0.5pt} \\
        \rule[2pt]{10pt}{0.5pt} \mathbf{v}_{\texttt{q[0]}} \rule[2pt]{10pt}{0.5pt} \\
        \vdots
    \end{bmatrix}}_{V \text{ (Actual Content)}}
    &=
    \underbrace{
    \begin{bmatrix}
        \rule[2pt]{10pt}{0.5pt} \mathbf{z}_{\texttt{rz}} \rule[2pt]{10pt}{0.5pt} \\
        \rule[2pt]{10pt}{0.5pt} \mathbf{z}_{\texttt{q[0]}} \rule[2pt]{10pt}{0.5pt} \\
        \vdots
    \end{bmatrix}}_{\text{Context Vectors}}
    \label{eq:MHSA}
\end{align*}

\subsubsection*{Sub-layer 2: Position-wise Feed-Forward Network (FFN)}
After the attention layer mixes information between tokens, this FFN processes each token's vector \textit{independently}. It is a simple two-layer neural network:
$$
\text{FFN}(z) = \max(0, zW_1 + b_1)W_2 + b_2
$$
This adds computational depth and allows the model to think about the new, context-rich information it just received from the attention sub-layer.

\subsubsection*{Residuals and Layer Normalization}
Each of these two sub-layers is wrapped by a \textbf{residual connection} and \textbf{layer normalization}.
\begin{itemize}
    \item \textbf{Residual Connection:} We take the input to the sub-layer and add it directly to the output: $\text{output} = \text{input} + \text{Sublayer}(\text{input})$. This skip-connection is vital for training deep networks, as it allows the original signal to propagate easily, preventing the model from forgetting the original input.
    \item \textbf{Layer Normalization:} This is a simple operation that re-centers and re-scales the vectors to have a mean of 0 and variance of 1. It acts as a regulator to keep the numbers from exploding or vanishing during training, which dramatically stabilizes the process.
\end{itemize}


\subsection*{The Decoder Stack}

The decoder's job is to generate the target sequence $\mathbf{y} = (y_1, \dots, y_M)$ token by token. It is also a stack of $N_x$ layers, which are very similar to the encoder's but with one crucial new sub-layer.

\begin{enumerate}
    \item \textbf{Masked Multi-Head Self-Attention:} The decoder's first sub-layer is a self-attention mechanism, just like in the encoder. However, when generating the $i$-th token, the decoder should only be allowed to see the tokens it has *already* generated (positions $1$ to $i-1$). To enforce this, we apply a look-ahead mask that sets all scores for future tokens to $-\infty$, effectively hiding them from the softmax.

    \item \textbf{Multi-Head Cross-Attention:} This is the layer that connects the encoder and decoder. It works just like self-attention, but its $Q, K, V$ vectors come from different places:
    \begin{itemize}
        \item \textbf{Queries ($Q$):} Come from the decoder's previous (masked) sub-layer. This represents what the decoder is trying to write next.
        \item \textbf{Keys ($K$) and Values ($V$):} Come from the final output $\mathbf{z}$ of the \textbf{encoder stack}. This represents the "full context of the entire input circuit."
    \end{itemize}
    This layer is where the model learns to map the input to the output. The decoder's query ("What should I write?") is matched against the encoder's keys ("Here's the info from the input circuit") to retrieve the values ("Here's the relevant part of the input circuit to pay attention to").

    \item \textbf{Position-wise Feed-Forward Network:} Identical to the encoder's FFN, this sub-layer processes each token's vector independently to think about the information it just gathered from both its own past (sub-layer 1) and the encoder's input (sub-layer 2).
\end{enumerate}
All these sub-layers also use residual connections and layer normalization.

\subsection*{Final Output and Sampling Strategies}

After the final decoder layer, we have a vector. This vector is passed through a final \textbf{Linear Layer} that projects it to a large vector (called logits) with a size equal to our entire target vocabulary. A \textbf{Softmax Function} is then applied to turn this vector into a probability distribution, where each token is assigned a probability of being the correct next token.

Once we have this probability distribution, we must decide how to select the next token. This is not always a simple choice, as the best token (highest probability) may not lead to the best overall sequence.

\subsubsection*{Greedy Search}
This is the simplest strategy. At each step, we simply select the single token with the highest probability.
\begin{itemize}
    \item \textbf{Pro:} It is fast, efficient, and deterministic. For a task like transpilation where there is often one correct answer, it is a very strong baseline.
    \item \textbf{Con:} It can be short-sighted. It might pick a token that looks good *now* but traps it in a sub-optimal sequence later. It is also prone to getting stuck in repetitive loops.
\end{itemize}

\subsubsection*{Temperature}
Temperature is a parameter $T$ (where $T > 0$) that is applied to the logits \textit{before} the softmax function. The logits are divided by $T$.
$$
\text{Probability}(token_i) = \text{softmax}\left(\frac{\text{logit}_i}{T}\right)
$$
\begin{itemize}
    \item \textbf{$T \approx 1$:} The distribution is unchanged.
    \item \textbf{$T < 1$ (e.g., 0.7):} The distribution becomes sharper or peakier. The model becomes more confident and conservative, increasing the probability of the most likely tokens and decreasing the probability of unlikely ones. This makes it more like greedy search.
    \item \textbf{$T > 1$ (e.g., 1.2):} The distribution becomes flatter. The model becomes more creative or uncertain, giving more probability mass to less likely tokens. This increases diversity but also the risk of errors.
\end{itemize}
After applying temperature, we sample from the new distribution.

\subsubsection*{Top-K Sampling}
Instead of considering all tokens in the vocabulary, we first identify the $K$ tokens with the highest probabilities. We then discard all other tokens, redistribute the probability mass among just these $K$ tokens, and then sample from this new, smaller distribution.
\begin{itemize}
    \item \textbf{Pro:} It prevents the model from picking absurdly unlikely tokens that might exist in the long tail of the probability distribution, which can happen with high-temperature sampling.
    \item \textbf{Con:} The number $K$ is fixed. If the model is very confident and the probability is concentrated in 3 tokens, a $K=50$ setting is useless. If the model is uncertain and the probability is spread across 100 tokens, $K=50$ might cut off good candidates.
\end{itemize}

\subsubsection*{Top-P (Nucleus) Sampling}
This is a more dynamic approach. Instead of picking a fixed number $K$, we pick a cumulative probability $P$ (e.g., $P=0.95$). We sort the tokens from most to least probable and sum their probabilities until we reach $P$. This nucleus of tokens is kept, all others are discarded, and we sample from the nucleus.
\begin{itemize}
    \item \textbf{Pro:} The size of the nucleus is dynamic. If the model is confident, the nucleus might only contain 2-3 tokens. If the model is uncertain, the nucleus might contain 50+ tokens. This adapts to the model's own confidence at each step.
    \item \textbf{Con:} It is more computationally complex than Top-K.
\end{itemize}
For our task, since transpilation requires high precision, a deterministic (Greedy) or near-deterministic (low temperature, or Top-P with a low $P$) strategy is generally preferred over highly creative sampling.

\section{\label{app:snippet_example}From QASM to tokens, then back to QASM}

\begin{figure}[H]
    \centering
    \includegraphics[width=0.65\textwidth]{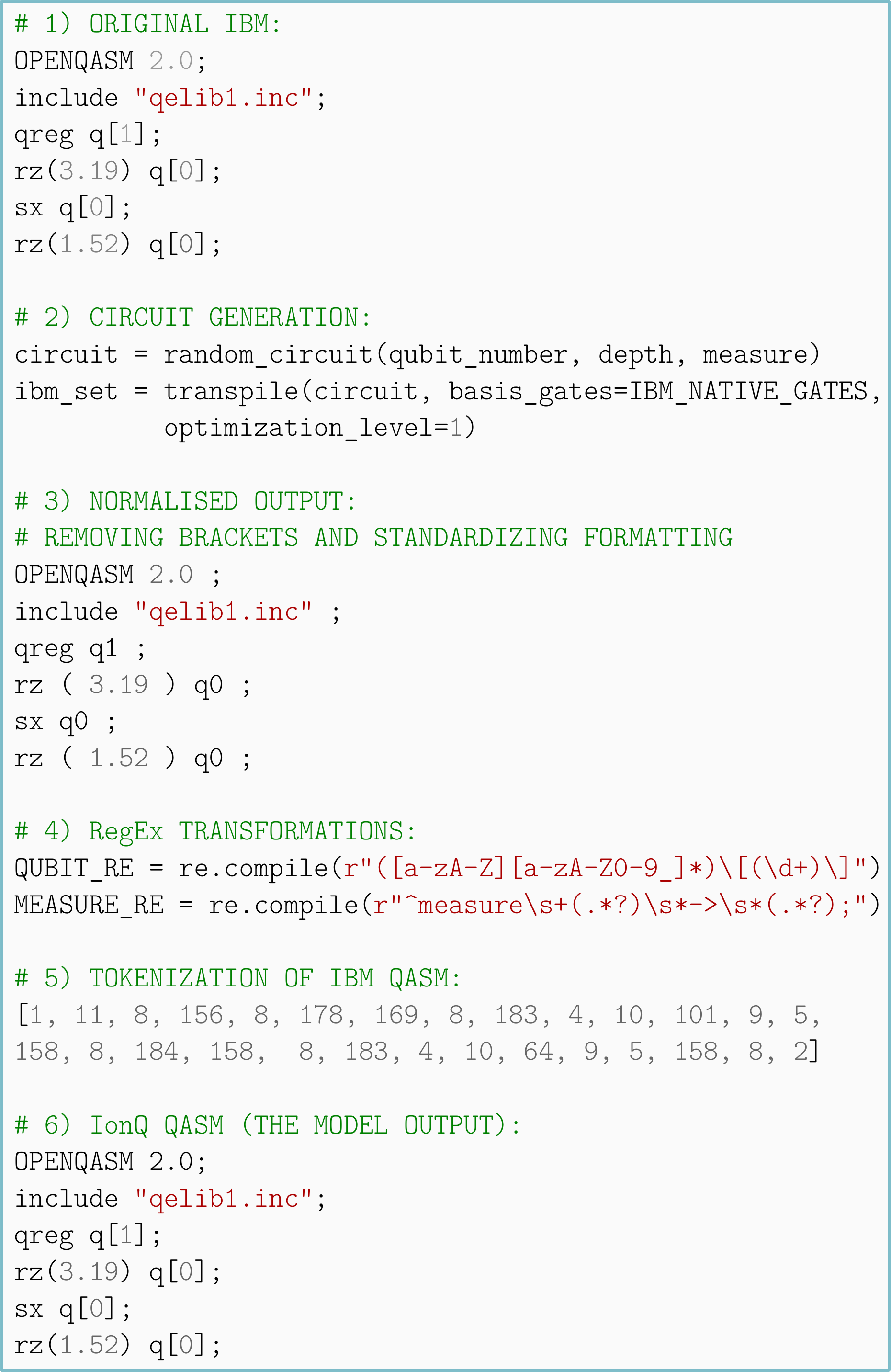}
    \caption{Snippet codes showing the evolution of the QASM input files throughout the encoding into tokens and back to a QASM format.}
    \label{fig:QASMelaboration}
\end{figure}

The purpose of this Appendix is to outline how to convert the QASM standard to its tokenized translation, and then its decoding back to a QASM representation. We are showing this pattern through a sequence of snippet codes, which indeed remarks the pipeline highlighted in Figure \ref{fig:pipeline}. In Figure \ref{fig:QASMelaboration}, we showcase how a QASM file is encoded into a sequence of tokens, fed into the transformer model and thereafter converted back into another QASM file suitable for the target platform (in our case, IonQ). In the first place, we start from a QASM file deploying the IBM native gates, as in the code \# 1) from Figure \ref{fig:QASMelaboration}. The QASM file is generated by a sequence of qiskit commands, which return a random circuit decomposed into the native gates of IBM. These commands are reported in the snippet \# 2) from Figure \ref{fig:QASMelaboration}.
Thereafter, the above text is simplified removing its redundant expressions with no grammatical meaning, e.g. the square brackets. The tool which allows to filter away such a redundancy of components is the RegEx, already cited in Section \ref{tokenizer}. The output turns to be as in the snippet \# 3) from Figure \ref{fig:QASMelaboration}.
The commands to invoke the RegEx, instead, are given by \# 4) in Figure \ref{fig:QASMelaboration}.
Thus, the next step is to convert the (simplified) QASM text to a proper numerical encoding for the transformers, i.e. the tokens. The tokenization directly converts the QASM file into a sequence of numbers. The tokenized representation of the QASM example is under the snippet \# 5).
This vector is directly fed into the transformer architecture, which in turn yields another vector of tokens representing the new output QASM file. Translating back the output tokens into this QASM file, the result we achieve in its final form is exposed as \# 6).
This sequence of commands from the IonQ QASM can be promptly converted into a circuit by the qiskit compiler, in order to evaluate its corresponding unitary and its fidelity with respect to the unitary from the original IBM circuit.

\section{\label{app:gates}Native gates}

In this Section, we provide the algebraic representation of the native gates for IBM and IonQ. The information can be checked on \href{https://quantum.cloud.ibm.com/docs/en/api/qiskit}{Qiskit API}.
\newline
\newline
\textbf{IBM} The native gates for IBM include, in the first place, the $\hat X$ Pauli matrix:
\begin{equation}
    \hat X = \begin{pmatrix}
        0 & 1 \\
        1 & 0
    \end{pmatrix}
\end{equation}
A correlated gate is the square root of $\hat X$, $\hat{SX} = \sqrt{\hat X}$:
\begin{equation}
    \hat{SX} = \frac{1}{2} \begin{pmatrix}
        1+i & 1-i \\
        1-i & 1+i
    \end{pmatrix}
\end{equation}
The rotation around the $z$ axis is defined as
\begin{equation}
    \hat R_z(\phi) = \begin{pmatrix}
        e^{-i\frac{\phi}{2}} & 0 \\
        0 & e^{i\frac{\phi}{2}}
    \end{pmatrix}
\end{equation}
At least, we have a two-qubits entangling gate, the CNOT gate:
\begin{equation}
    \hat{CX} = \begin{pmatrix}
        1 & 0 & 0 & 0 \\
        0 & 1 & 0 & 0 \\
        0 & 0 & 0 & 1 \\
        0 & 0 & 1 & 0
    \end{pmatrix}
\end{equation}
\newline
\textbf{IonQ} The IonQ machines account the rotations along all the axes:
\begin{equation}
    \hat R_x(\theta) = \begin{pmatrix}
        \cos(\frac{\theta}{2}) & i\sin(\frac{\theta}{2}) \\
        i\sin(\frac{\theta}{2}) & \cos(\frac{\theta}{2}) 
    \end{pmatrix}, \quad
    \hat R_y(\theta) = \begin{pmatrix}
        \cos(\theta) & \sin(\theta) \\
        -\sin(\theta)& \cos(\theta)
    \end{pmatrix}, \quad
    \hat R_z(\theta) = \begin{pmatrix}
        e^{-i\frac{\theta}{2}} & 0 \\
        0 & e^{i\frac{\theta}{2}}
    \end{pmatrix}
\end{equation}
As entangling gate, IonQ provides the double rotation (i.e. involving two qubits) around the $x$ axis:
\begin{equation}
    \hat R_{xx}(\theta) = \begin{pmatrix}
        \cos(\frac{\theta}{2}) & 0 & 0 & -i\sin(\frac{\theta}{2}) \\
        0 & -i\sin(\frac{\theta}{2}) & \cos(\frac{\theta}{2}) & 0 \\
        0 & \cos(\frac{\theta}{2}) & -i\sin(\frac{\theta}{2}) & 0 \\
        -i\sin(\frac{\theta}{2}) & 0 & 0 & \cos(\frac{\theta}{2}) 
    \end{pmatrix}
\end{equation}

\section{\label{app:SKscalingLaws}The Solovay-Kitaev scaling laws for the tokenization process}

The Solovay-Kitaev theorem states that, once a discrete subset $\mathcal{G} \subset SU(2)$ of logic gates, dense in $SU(2)$, is given, any single-qubit unitary $U$ can be approximated with precision $\epsilon$ in $O(\log^c(1/\epsilon))$ gates from $\mathcal{G}$~\cite{nielsen2010quantum,dawson2005solovay,maronese2022quantum}. The constant $c$ is assessed to range between $2$ and $4$, depending on the formulation of the theorem~\cite{dawson2005solovay}.
This theorem plays a crucial role in quantum compiling theory, since it states that, in order to implement a fault-tolerant gate $U(\theta)$ through a set of native gates, it takes only a polylogarithmic number of gates, avoiding worse scaling (such as a polynomial or even an exponential one).
The Solovay-Kitaev theorem allows to convert any circuit in a discrete set of fault-tolerant gates -- for instance, the Clifford gates and the $S$ one -- at the cost of introducing a certain degree of precision $\epsilon$ when approximating the single-qubit rotations~\cite{dawson2005solovay}.
Moreover, the theorem do not need to address a specific universal set of gates, since any of universal set fits the hypothesis of the theorem. Such a statement means that, once a specific hardware display its own universal set of native logic gates, the Solovay-Kitaev theorem can be applied via such a suitable set~\cite{maronese2022quantum}. In other words, the Solovay-Kitaev theorem is agnostic about the hardware and the platform we are considering.

Consider now a circuit with length $m$, $\hat U_1 \circ \hat U_2 \circ ... \circ \hat U_m$, which we want to represent with precision $\varepsilon$. Thus, a precision $\varepsilon/m$ is required for each single-qubit gate to be decomposed. This argument descends from the definition of distance in the unitary group $SU(2)$, $\|U, V\| = \max_{\|\psi\|=1} \|(U-V)\ket{\psi}\|$, so that, comparing two sequences $U_1 U_2$ and $V_1 V_2$ we may exploit the triangular inequality~\cite[pag.195]{nielsen2010quantum}:
\begin{equation}
    \| U_1 U_2 - V_1 V_2 \| = \| U_1 U_2 - U_1 V_2 + U_1 V_2 - V_1 V_2 \| \leq \| U_1 (U_2 - V_2) \| + \| (U_1 - V_1) V_2 \| = \| U_2 - V_2 \| + \| U_1 - V_1 \|
\end{equation}
Since $U_i$ and $V_i$ can be set as sequences of operators, the same argument can be repeated iteratively.
The overall length of the circuit, therefore, scales as follows:
\begin{equation}
\label{eq:SKscaling}
    O\left(m \log^c\left(\frac{m}{\varepsilon}\right)\right)
\end{equation}
with $m$ being of course the length of the circuit, and the logarithm the length of the decomposition of the single-qubit gates. 

We want now to determine how the complexity of the model does scale when translating a circuit decomposed via the Solovay-Kitaev theorem from a set $\mathcal{G}_1$, suitable for a specific hardware, to a $\mathcal{G}_2$ one for another platform. In the first place, we have observed that the complexity of the model scales as $N  m$, with $N$ being the number of tokens required and $m$ the length of the circuit. Therefore, the number of neurons required to achieve a specific precision $\varepsilon$ for the entire circuit is $\varepsilon/mN$. An important consideration arise, since we do not need anymore to sample a certain number of angles from the range $[0, 2\pi]$. In fact, the rotations $\hat U(\theta)$ are replaced by sequences of elements from $\mathcal{G}$. The number of tokens required, thus, reduces to $N_g = N_{source} + N_{target}$ -- with $N_{source}$ and $N_{target}$ being, respectively, the cardinality of $\mathcal{G}_1$ and $\mathcal{G}_2$ -- and number $n$ of qubits involved. 
Now, instead of considering a generic sequence $m$ fixed, let's study an algorithm whose number of gates scales as $f(n)$, with $n$ being the number of qubits in the register. We can easily see that $N_q + N_{g} \sim N_q$, since the number of available gates is fixed by the hardware we are considering.
From such a consideration, we can state that the complexity of the model, once the quantum circuit has been decomposed though the Solovay-Kitaev, would scale as
\begin{equation}
\label{eq:SKTransformerscaling}
    O\left(nf(n) \log^c\left(\frac{n f(n)}{\varepsilon}\right)\right)
\end{equation}

\end{document}